\documentclass[12pt,letterpaper]{article}

\usepackage{times}
\usepackage[sort&compress]{natbib}
\setcitestyle{super,open={},close={},citesep={,}}
\def\Colored{1}

\renewcommand\refname{References}

\makeatletter
\def\@biblabel#1{#1.}
\makeatother

\usepackage{graphicx}
\graphicspath{{./FigureMain/}{./FigureSupp/}}

\usepackage{amsmath,amssymb}
\usepackage{bm}

\usepackage{here}

\usepackage{color}
\usepackage{ulem}

\ifodd \Colored
  
  \newcommand{\delete}[1]{\textcolor{blue}{\sout{#1}}}
\else
  
  \newcommand{\delete}[1]{{}}
\fi

\makeatletter
 \long\def\@makecaption#1#2{%
 \vskip\abovecaptionskip
\sbox\@tempboxa{\textbf{#1 $\vert$ }\ #2}%
 \ifdim \wd\@tempboxa >\hsize
\textbf{#1 $\vert$ }\ #2\relax\par
 \else
 \fi
 \vskip\belowcaptionskip}
 \makeatother

\newcommand\figref[1]{Fig.~\ref{fig:#1}}

\newcommand\figureref[1]{Figure~\ref{fig:#1}}

\makeatletter
\@ifundefined{textcolor}{}
{%
 \definecolor{BLACK}{gray}{0}
 \definecolor{WHITE}{gray}{1}
 \definecolor{RED}{rgb}{1,0,0}
 \definecolor{GREEN}{rgb}{0,.4,0}
 \definecolor{BLUE}{rgb}{0,0,1}
 \definecolor{CYAN}{cmyk}{1,0,0,0}
 \definecolor{MAGENTA}{cmyk}{0,1,0,0}
 \definecolor{YELLOW}{cmyk}{0,0,1,0}
 }
\makeatother

\newcommand{\blk}{\protect\color{BLACK}}

\def\op#1{\hat{#1}}
\def\opvec#1{\op{\vec{#1}}}

\def\id{I}

\def\mat#1{\bm{\mathrm{#1}}}
\renewcommand{\vec}[1]{\bm{\mathrm{#1}}}

\def\tp{\mathrm{T}}

\DeclareMathOperator{\sech}{sech}

\providecommand{\ket}[1]{\left\vert{#1}\right\rangle}

\def\negspace{\!}

\def\lsub#1#2{{\protect\vphantom{#1}}_{#2} \negspace {#1}}

\def\inprod#1#2{\left\langle {#1} | {#2} \right\rangle}

\title{
Ultra-Large-Scale Continuous-Variable Cluster States Multiplexed in the Time Domain
}

\author
{Shota Yokoyama$^{1}$, Ryuji Ukai$^{1}$, Seiji C.\ Armstrong$^{1,2}$,\\
Chanond Sornphiphatphong$^{1}$, Toshiyuki Kaji$^{1}$, Shigenari Suzuki$^{1}$,\\
Jun-ichi Yoshikawa$^{1}$, Hidehiro Yonezawa$^{1}$,\\
Nicolas C.\ Menicucci$^{3}$, Akira Furusawa$^{1\ast}$\\
\\
\normalsize{$^{1}$Department of Applied Physics, School of Engineering, The University of Tokyo,}\\
\normalsize{7-3-1 Hongo, Bunkyo-ku, Tokyo 113-8656, Japan}\\
\normalsize{$^{2}$Centre for Quantum Computation and Communication Technology,}\\
\normalsize{Department of Quantum Science, The Australian National University,}\\
\normalsize{Canberra, ACT 0200, Australia}\\
\normalsize{$^{3}$School of Physics, The University of Sydney, NSW, 2006, Australia}\\
\\
\normalsize{$^\ast$To whom correspondence should be addressed; E-mail: akiraf@ap.t.u-tokyo.ac.jp}
}

\date{\empty}

\begin{document} 


\baselineskip24pt


\maketitle

\clearpage

{\bf 
\noindent
Quantum computers promise ultrafast performance of certain tasks \cite{Nielsen00}. Experimentally appealing, measurement-based quantum computation~(MBQC) \cite{Raussendorf01} requires an entangled resource called a cluster state \cite{Briegel2001}, with long computations requiring large cluster states.
Previously, the largest cluster state consisted of 8 photonic qubits \cite{Yao12} or light modes \cite{Su12}, while the largest multipartite entangled state of any sort involved 14 trapped ions \cite{Monz11}. 
These implementations involve quantum entities separated in space, and in general, each experimental apparatus is used only once. Here, we circumvent this inherent inefficiency by multiplexing light modes in the time domain. 
We deterministically generate and fully characterise a continuous-variable cluster state \cite{Zhang06,Menicucci06} containing more than 10,000 entangled modes. 
This is, by 3 orders of magnitude, the largest entangled state ever created to date.
The entangled modes are individually addressable wavepackets of light in two beams.
Furthermore, we present an efficient scheme for MBQC on this cluster state based on sequential applications of quantum teleportation.
}\\

Originally formulated as a demonstration as to why quantum mechanics must be incomplete in the famous 1935 Einstein-Podolsky-Rosen (EPR) paradox \cite{Einstein35}, entanglement is now recognized as a signature feature of quantum physics \cite{Bell1964}, and it plays a central role in various quantum information processing (QIP) protocols \cite{Nielsen00,Furusawa11}.
For example, the bipartite entangled state known as an EPR state \cite{Einstein35} is a resource for quantum teleportation (QT), whereby a quantum state is transferred from one location to another without physical transfer of the quantum information \cite{Bennett93,Furusawa98,Lee11}.

Measurement-based quantum computation (MBQC) \cite{Raussendorf01,Zhang06,Menicucci06,Walther05,Tokunaga08,Ukai11-1,Ukai11-2}, which is based on the QT of information and logic gates, requires the special class of multipartite entangled resource states known as cluster states~\cite{Briegel2001}. The number of entangled quantum entities and their entanglement structure (represented by a graph) determines the resource space available for computation. Ultra-large-scale QIP (which could be based on MBQC) will require ultra-large-scale entangled states \cite{Raussendorf01,Zhang06,Menicucci06}.

In the vast majority of optical experiments, quantum modes are distinguished from each other by their spatial location. This leads to an inherent lack of scalability as each additional entangled party requires an increase in laboratory equipment and dramatically increases the complexity of the optical network \cite{Yukawa08,Peter07}. Further, due to the probabilistic nature of photon pair generation, demonstrations involving the postselection of photonic qubits\cite{Yao12,Walther05,Tokunaga08} suffer from dramatically reduced event success rates with each additional qubit. 

One method to overcome this problem of scalability is to deterministically encode the modes within one beam. Entanglement between quadrature-phase amplitudes in continuous-wave laser beams has been deterministically created and exploited in QIP \cite{Su12,Furusawa98,Lee11,Ukai11-1,Ukai11-2,Yukawa08,Seiji12,Pysher11,Roslund13}, even though the quantum correlations are finite. Previous attempts to deterministically create cluster states within one beam have exploited the spatial \cite{Seiji12} or spectral \cite{Pysher11,Roslund13,Menicucci08} orthogonality of quantum modes. 
While such methods are potentially scalable, current experimental results are limited to generating entangled states of just a few modes each \cite{Seiji12,Pysher11,Roslund13}.\setcitestyle{numbers,open={},close={},citesep={,}}A novel method proposed in ref.~\cite{Menicucci11} lets quantum modes propagate within the same beam --- distinguished and ultimately made orthogonal by their separation in time.\setcitestyle{super,open={},close={},citesep={,}}The time-domain multiplexing approach allows each additional quantum mode to be manipulated by the same optical components at different times, which is a powerful concept, as found in atomic ensemble quantum memories \cite{Usmani2010}. 

Here, we demonstrate the deterministic generation of ultra-large-scale entangled states consisting of more than $10,000$ entangled wave-packets of light, multiplexed in the time domain.
The generated states, which we call extended EPR (XEPR) states, are equivalent up to local phase shifts to topologically one-dimensional continuous-variable (CV) cluster states \cite{Menicucci11} and are therefore a universal resource for single-mode MBQC with continuous-variables \cite{Menicucci06}. Fully universal multimode MBQC is achievable simply by combining two XEPR states with differing time delays on two additional beam-splitters \cite{Menicucci11}. 
Note that in our time-domain multiplexed demonstration only a small part of the entire XEPR state exists at each instant of time. This is the reason why we could achieve such an efficient setup for the creation and verification of an ultra-large number of entangled modes.

\begin{figure}[p]
	\centering
	\includegraphics[scale=1,clip]{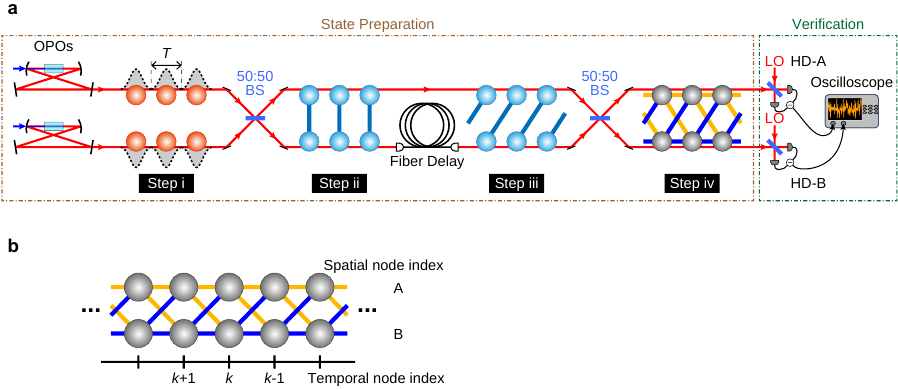}
	\caption{
\noindent {\bf 
Schematic of the experimental setup and ultra-large-scale XEPR state.}\newline 
\textbf{a}, Generation of ultra-large-scale cluster-state entanglement. Squeezed temporal modes are mixed on a beam-splitter to create EPR states, which are then staggered by a fiber delay line. The staggered EPR-state wave-packets are then mixed to generate a continuous graph structure, where each wave-packet is entangled to neighbouring wave-packets. Nodes (coloured spheres) and links between them represent optical wave-packets and entanglement, respectively.
Independent quantum states exist in each temporally localised wave-packet with time duration $T$. 
OPO, optical parametric oscillator; 
50:50 BS, balanced beam-splitter; 
HD, homodyne detector; 
LO, local oscillator.
\textbf{b}, The extended EPR state, equivalent to a CV cluster state up to local phase shifts (redefinition of quadrature operators). See supplementary section `S2' for full details and graph representation.
}
	\label{fig:ExperimentalSetups}
\end{figure}

The XEPR states are generated by entangling together sequentially propagating EPR states contained within two beams. This can be viewed as four distinct steps, as illustrated in \figref{ExperimentalSetups}a.
First, two continuous-wave squeezed light beams are generated from two optical parametric oscillators (OPOs) (Step i). We divide the squeezed light beams into time bins of time period $T$, where $1/T$ is sufficiently narrower than the bandwidths of the identical OPOs. Wave-packets of light in each of the time bins represent mutually independent (orthogonal) squeezed states.
Second, a series of EPR states separated by time interval $T$ are deterministically created by combining the two squeezed light beams on the first balanced beam-splitter (Step ii).
The quantum correlations that manifest from the beam-splitter interaction are represented by links between the nodes (coloured spheres). The nodes here represent the orthogonal wave-packets.
Third, the bottom-rail node of the EPR state is delayed for the duration $T$ after passing through a fiber delay line (Step iii). After the delay the top-rail node of each EPR state is synchronised in time with the bottom-rail node of the previous EPR state.
By combining the staggered EPR states on the second balanced beam-splitter, each EPR state interacts with the previous and successive EPR states (Step iv).
This leads to all of the wave-packets in each of the two rails being connected to neighbouring wave-packets by entanglement links, thereby producing the XEPR state. 

\begin{figure}
	\centering
	\hspace*{-2pc}
	\includegraphics[scale=1,clip]{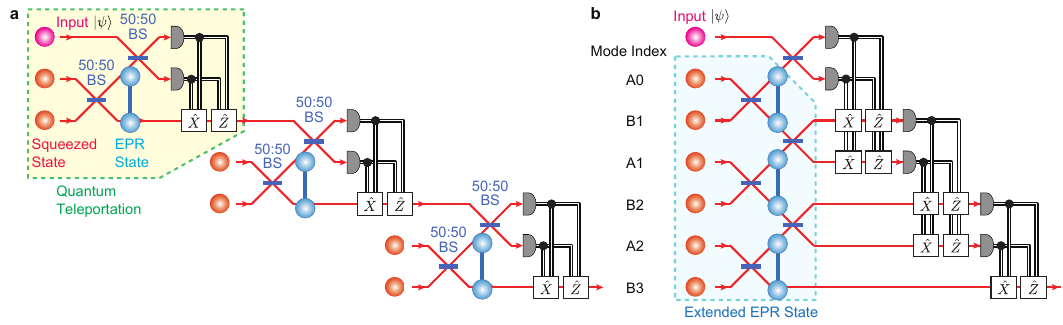}
	\caption{
\noindent {\bf 
XEPR states for sequential quantum teleportation.}\newline 
\textbf{a}, Circuit of standard sequential quantum teleportation. 
\textbf{b}, Circuit of sequential quantum teleportation through the XEPR state. 
50:50 BS, balanced beam-splitter; $\left|\psi\right\rangle$, arbitrary and unknown quantum state as the input of quantum teleporter; $\hat{X}$ or $\hat{Z}$, phase-space displacement operation based on measurement outcomes.
}
	\label{fig:Teleportation}
\end{figure}

The original proposal \cite{Menicucci11} for using the XEPR state as a resource state for MBQC is inefficient in its use of available squeezing resources. In \figref{Teleportation}, we introduce a new, efficient method based on sequential QT. 
QT in a strict sense is an identity operation on the input quantum state, but it can be generalized to unitary operations by appropriately changing the measurement bases, leading to MBQC. Each QT requires an EPR state and involves two single-mode measurements followed by subsequent phase-space displacement operations that depend on the measurement outcomes. The two circuits of \figureref{Teleportation}a and \figureref{Teleportation}b perform identical operations, due to the parallelism between reversible operations. The resource state in \figureref{Teleportation}b is precisely the XEPR state. 

Ideal quadrature-entangled states are simultaneous eigenstates of particular linear combinations of the quadrature operators, called nullifiers \cite{Gu2009,Menicucci2011}.
For example an ideal EPR state $\left| \mathrm{EPR}\right\rangle$, which is the ideal state approximated by the Gaussian state at step ii in \figref{ExperimentalSetups}a, is specified by the following nullifier relations:
\begin{align}
\label{EPR}
\left( \hat x^A-\hat x^B\right) \left| \mathrm{EPR}\right\rangle
&=0, \nonumber \\
\left( \hat p^A+\hat p^B\right) \left| \mathrm{EPR}\right\rangle
&=0.
\end{align}
Here, superscripts $A$ and $B$ denote two wave-packets. In our setup, they refer to the top rail and bottom-rail wave-packets, respectively.
$\hat x^Q$ and $\hat p^Q$ are the quadrature operators of a wave-packet $Q$, which do not commute for the same wave-packet: $\bigl[ \hat x^Q,\hat p^{R}\bigr] =$ $i\delta_{Q,R}/2$, where $\delta_{Q,R}$ is the Kronecker delta, and $\hbar$ is normalised to $1/2$. 
While $\hat x$ and $\hat p$ of a single wave-packet cannot be determined simultaneously due to the Heisenberg uncertainty principle, equation~\eqref{EPR} shows that the correlations between the two wave-packets are perfectly determined. More precisely, the quadrature amplitudes $\hat x$ of the two wave-packets are perfectly correlated ($x^A=x^B$), and the amplitudes $\hat p$ are perfectly anticorrelated ($p^A=-p^B$). 

In its ideal form, the XEPR state $\left|\mathrm{XEPR}\right\rangle$ generated in our experiment (\figref{ExperimentalSetups}b) is specified by 
\begin{align}
\label{graph}
\left( \hat x^A_k+\hat x^B_k+\hat x^A_{k+1}-\hat x^B_{k+1}\right) \left| \mathrm{XEPR}\right\rangle
&=0, \nonumber \\
\left( \hat p^A_k+\hat p^B_k-\hat p^A_{k+1}+\hat p^B_{k+1}\right) \left| \mathrm{XEPR}\right\rangle
&=0. 
\end{align} 
Here, $A$ and $B$ denote the two independent beams (top and bottom rail, respectively), while $k=1,2,\dots$ represents the temporal index.  
We consider the XEPR state to be a natural extension of EPR states because the nullifiers are composed of either $\hat x$ or $\hat p$ quadratures, but not both. 
We see that the addition of two quadratures $\hat x^A_k+\hat x^B_k$ (respectively, $\hat p^A_k+\hat p^B_k$) at any given time index $k$ have negative (positive) correlations with the difference of the two quadratures $\hat x^A_{k+1}-\hat x^B_{k+1}$ (respectively, $\hat p^A_{k+1}-\hat p^B_{k+1}$) at the subsequent time index ${k+1}$.
The XEPR state is entirely equivalent to a CV cluster state under an appropriate redefinition of quadrature operators ($\hat{x}\to\hat{p}$, $\hat{p}\to-\hat{x}$) for every other temporal mode \cite{Menicucci2011}. Such a redefinition is completely passive and does not change the resource requirements for MBQC using this state, but we choose to call it an XEPR state rather than a CV cluster state because our verification procedure takes advantage on the fact that the ideal nullifiers only involves $\hat{x}$s and $\hat{p}$s.

In reality, the generated XEPR states --- and therefore the nullifiers --- have excess noise due to the unphysical nature of infinite squeezing. 
Despite this, the full inseparability of the state can be shown when the resource squeezing level is high enough.
The sufficient conditions for fully inseparable entanglement \cite{Peter03} are given by the variances as
\begin{align}
\label{Inseparability}
\bigl\langle \left( \hat x^A_k+\hat x^B_k+\hat x^A_{k+1}-\hat x^B_{k+1}\right)^2\bigr\rangle 
&<\frac{1}{2} \nonumber \\
\text{and} \quad \bigl\langle \left( \hat p^A_k+\hat p^B_k-\hat p^A_{k+1}+\hat p^B_{k+1}\right)^2\bigr\rangle 
&<\frac{1}{2}
\end{align}
for all $k$, where a bracket $\hat{O}\ (\langle\hat{O}\rangle)$ denotes the expectation value of an operator $\hat O$.

\begin{figure*}
	\centering
	\includegraphics[scale=1,clip]{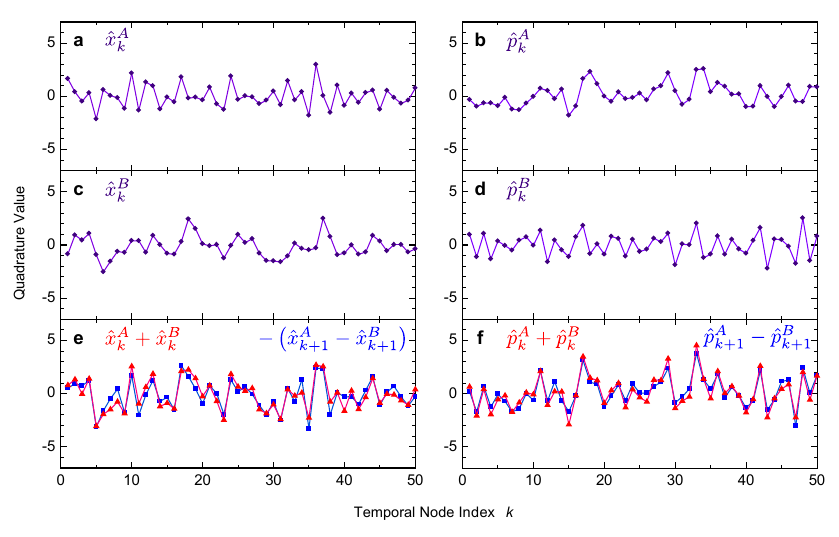}
	\caption{
\noindent {\bf 
Measured quantum correlations of the first $50$ wave-packets.}\newline 
\textbf{a}--\textbf{d}, Measured quadrature amplitudes of $\hat x^A_k$, $\hat p^A_k$, $\hat x^B_k$ and $\hat p^B_k$ of temporally localised wave-packets
at temporal node positions $k$. They are randomly distributed around zero. 
\textbf{e},\textbf{f}, Additions and time-shifted subtractions of certain quadratures show quantum correlations displayed by the near perfect overlaps.
}
	\label{fig:Quadrature}
\end{figure*}

The experimental quadrature amplitudes $\hat x^A_k$, $\hat p^A_k$, $\hat x^B_k$ and $\hat p^B_k$ of the first $50$ wave-packets are plotted in \figref{Quadrature}(a--d).
We see they are randomly distributed around zero. Linear combinations of the quadrature amplitudes at neighbouring times exhibit quantum correlations as in  equation~\eqref{graph} and are shown in \figref{Quadrature}(e,f).
The amplitudes almost perfectly overlap, showing strong anti-correlations and correlations in both quadrature combinations.

\begin{figure*}
	\centering
	\includegraphics[scale=1,clip]{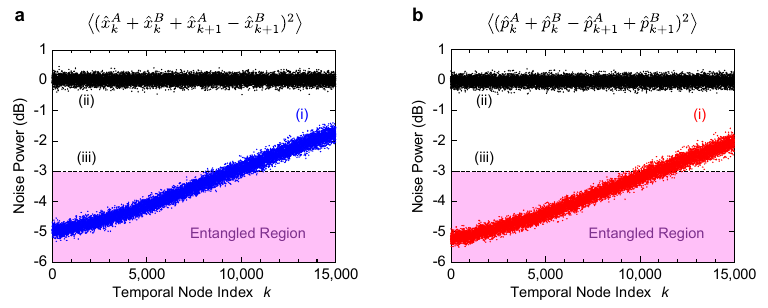}
	\caption{
\noindent {\bf 
Quantum correlations of the XEPR state for the first $30,000$ wave-packets.}\newline 
\textbf{a,b}, (i) Measured nullifier variances for the first $15,000$ temporal mode positions ($30,000$ wave-packets in both rails) are shown for the $\hat x$ quadrature in blue (\textbf{a}) and for the $\hat p$ quadrature in red (\textbf{b}), as in equation~\eqref{Inseparability}. (ii) measured variances for vacuum inputs, used as a reference for noise power. (iii) $-3.0$~dB lines indicate the bounds of inseparability; data points below this line demonstrate entanglement.
}
	\label{fig:Variance}
\end{figure*}

In order to quantify the quantum correlations, we repeat the single-shot generation of the entire XEPR state more than $3,000$ times, allowing us to measure the variances at each temporal position. We then evaluate the multi-partite inseparability criteria given in equation~\eqref{Inseparability}. The measurement results are shown in \figref{Variance}. The variances for the XEPR states are shown by traces (i). The bound of inseparability given in equation~\eqref{Inseparability} corresponds to $-3.0$ dB, shown by the dashed lines (iii). We see that the experimental XEPR state is clearly below the bound of inseparability in the entangled region for more than $5,000$ temporal positions. Given the dual-rail structure of the XEPR state, the two beams $A$ and $B$ each contain the same number of wave-packets, so that $5,000$ temporal positions corresponds to $10,000$ wave-packets.
Vacuum-state inputs are used as references, and traces around $0$~dB (ii) show the variances of the nullifiers for the vacuum states. 

The mean variances of the first $1,000$ points in the $\hat x$ and $\hat p$ quadratures are 
$-4.9\pm 0.2$ dB
and
$-5.2\pm 0.2$ dB,
respectively.
Note that absolutely no corrections for any losses are performed.

The variances of the XEPR state steadily increase (and therefore the entanglement degrades) with time for technical reasons related to our control scheme.
During the data acquisition process we switch off all active feedback control of the optical setup. This is in order to avoid any unwanted noise arising from the feedback that will degrade our measurements. The increase in variance is therefore explained by the relative phase drifts of the entangled state caused by disturbances from the environment.

In summary, we have experimentally demonstrated the generation of ultra-large-scale entangled states in a deterministic fashion. More than $10,000$ wave-packets of light are shown to be fully inseparable 
in a CV cluster-state configuration. Fault-tolerant quantum computation will additionally require efficient encoding and error correction \cite{Gu2009}. Compared to the largest entangled states previously engineered, the number of entangled modes here is larger by three orders of magnitude. Due to their sheer size, regular structure, and deterministic method of creation, we fully expect that these ultra-large-scale states will enable other QIP applications in addition to MBQC.

\vspace{\baselineskip}

\noindent
{\bf\large Methods}\\
The squeezing levels of our OPOs were $-6$~dB with a bandwidth of $34$~MHz.
The optical fiber length was $30$~m, corresponding to the time duration of the wave-packets $T=157.5$~ns. Homodyne detection is employed to measure the quadrature amplitudes of each wave-packet. The signals of the homodyne detectors are integrated with the non-overlapping temporally-localised mode functions of the wave-packets.

\paragraph*{Acknowledgments:}
This work was partly supported by Project for Developing Innovation Systems (PDIS), Grants-in-Aid for Scientific Research (GIA), Global Center of Excellence (G-COE), and Advanced Photon Science Alliance (APSA) commissioned by the Ministry of Education, Culture, Sports, Science and Technology (MEXT) of Japan, Funding Program for World-Leading Innovative R\&D on Science and Technology (FIRST) initiated by Council for Science and Technology Policy (CSTP) of Japan, and Australian Research Council (ARC) Centre of Excellence for Quantum Computation \& Communication Technology (CQC2T), project number CE110001027.
S.Y.\ acknowledges financial support from Advanced Leading Graduate Course for Photon Science (ALPS). 
R.U.\ acknowledges support from Japan Society for the promotion of Science (JSPS). 
S.A.\ acknowledges financial support from the Prime Minister's Australia Asia Award. 
N.C.M.\ was supported by the ARC under grant No.~DE120102204.

{\noindent}{\bf Author contributions}

{\noindent}S.Y., R.U.\ and S.A.\ planned and designed the experiment under the supervision of J.Y., H.Y.\ and A.F., based on the proposal by N.C.M\null. S.Y.\ designed the experimental setup. C.S, T.K.\ and S.Y.\ constructed the optical setup, and S.S.\ built the fiber alignment system. R.U., S.A., N.C.M.\ and J.Y.\ formulated the theory. R.U.\ designed and constructed the data acquisition system. S.A.\ designed and constructed the digital control system. R.U., S.A., T.K.\ and S.Y.\ conducted the data analysis. H.Y.\ assisted in noise analysis. S.Y.\ and S.A.\ wrote the manuscript with assistance from the team. 

{\noindent}
{\bf Additional information}

{\noindent}
Supplementary information is available in the online version of the paper.
Preprints and permissions information is available online at www.nature.com/reprints.
Correspondence and requests for materials should be addressed to A.F\null.

{\noindent}
{\bf Competing financial interests}

{\noindent}
The authors declare no competing financial interests.


 \clearpage

\clearpage

\renewcommand{\thesection}{S\arabic{section}}
\renewcommand{\thesubsection}{S\arabic{section}.\arabic{subsection}}
\renewcommand{\thesubsubsection}{S\arabic{section}.\arabic{subsection}.\arabic{subsubsection}}

\def\theequation{S\arabic{equation}}
\def\refname{Supplementary References}

\makeatletter
\renewcommand{\figurename}{Fig.}
\renewcommand{\thefigure}{S\arabic{figure}}

 \long\def\@makecaption#1#2{%
 \vskip\abovecaptionskip
 \sbox\@tempboxa{\textbf{#1.}\ #2}%
 \ifdim \wd\@tempboxa >\hsize
 \textbf{#1.}\ #2\relax\par
 \else
 \global \@minipagefalse
 \hbox to\hsize{\hfil\box\@tempboxa\hfil}%
 \fi
 \vskip\belowcaptionskip}
\makeatother

\makeatletter
\newcommand\fleqnoff{\@fleqnfalse\@mathmargin\@centering}
\newcommand\fleqnon[1][\leftmargini]{\@fleqntrue\@mathmargin=#1\relax
\@ifundefined{mathindent}{\let\mathindent\@mathmargin}{}}
\makeatother
\newcommand\Expbig[1]{\bigl\langle #1 \bigr\rangle}
\newcommand\Varbig[1]{\bigl\langle #1^2 \bigr\rangle}

\baselineskip24pt

\begin{center}
{\LARGE
Supplementary Information for \\
Ultra-Large-Scale Continuous-Variable Cluster States Multiplexed in the Time Domain}
\end{center}

\section{Experimental Setup}

\begin{figure}[!b]
	\centering
	\includegraphics[scale=0.5,clip]{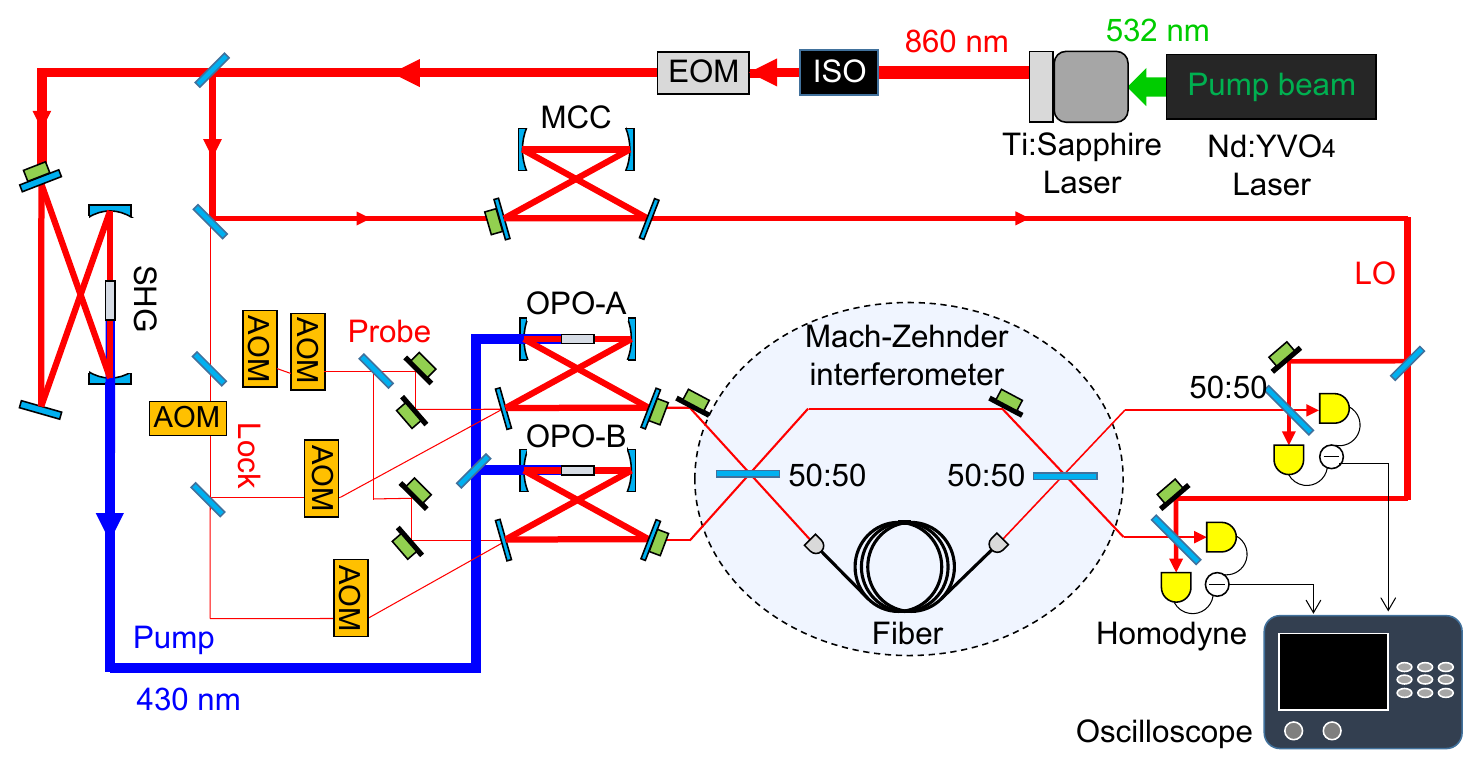}
	\caption{Experimental setup.}
	\label{fig:SupExperiment}
\end{figure}

\noindent
\figureref{SupExperiment} shows the schematic of the experimental setup.
In this setup, we employ a continuous-wave (CW) Ti:Sapphire laser (SolsTiS-SRX, M Squared Lasers) operating at $860$~nm as the primary optical source.
The pump laser is a laser-diode-pumped and frequency-doubled Nd:YVO$_4$ laser (Verdi V-10, Coherent).
10 W of the pump beam from the Nd:YVO$_4$ results in $1.9$~W output power of the Ti:Sapphire laser.

The 860~nm fundamental beam passes through the optical isolator (ISO; FI850-5SV, Linos) and the electro-optic phase modulator (EOM; PM25, Linos).
The phase modulation by the EOM adds 16.5~MHz sideband components which are utilised for locking all the optical cavities via the Pound-Drever-Hall locking technique.

Almost half of the fundamental beam is sent to a second harmonic generator (SHG) and converted to approximately $500$~mW of a $430$~nm beam.
The SHG is a bow-tie cavity consisting of two spherical mirrors (radius of curvature $50$~mm), two flat mirrors, and contains a $10$~mm $\times$ $3$~mm $\times$ $3$~mm KNbO$_3$ crystal.
Its round trip length is $500$~mm, and the input coupler transmissivity is $10\%$.
The rest of the fundamental beam is further split and distributed for controlling the sub-threshold optical parametric oscillators (OPOs), the interferometers, the homodyne detectors, and so on.

The second harmonic from the SHG is injected into the OPOs as their pump beam ($100$~mW for each OPO) to generate the squeezed vacuum beams, which are the quantum resources for this experiment.
The OPO is a bow-tie cavity consisting of two spherical mirrors (radius of curvature $38$~mm), two flat mirrors and a $10$~mm $\times$ $1$~mm $\times$ $1$~mm periodically poled KTiOPO$_4$ (PPKTP) crystal.
Its round trip length is $230$~mm, the half width at half maximum is $17$~MHz, the output coupler transmissivities are $14.9\%$ (OPO-A) and $15.4\%$ (OPO-B), and intra cavity losses are $0.41\%$ (OPO-A) and $0.30\%$ (OPO-B).
The OPO is locked via the Pound-Drever-Hall locking method by introducing a locking beam which is appropriately frequency-shifted and counter-propagating in the cavity so as to avoid interference with the squeezed vacuum beam.

The squeezed vacuum states are combined by a Mach-Zehnder interferometer (MZI) with asymmetric arm lengths. 
After the first balanced beam-splitter, they are converted into two-mode EPR states. The EPR states become staggered due to the asymmetric arm lengths. The staggered EPR states are then combined in sequence at the second balanced beam-splitter, forming the extended EPR state.

For the optical delay line to asymmetrise the MZI, we employ an optical fiber. 
To minimize the insertion loss of the delay line, we employ fiber patchcodes with special anti-reflection (AR) coatings at $860$~nm on both ends (PMJ-3A3A-850-5/125-1-2-1-AR2, OZ Optics).
We fabricated arbitrary lengths of fiber cables by splicing the patchcodes and bare fibers (SM85-PS-U40A, Fujikura).
The lenses for focusing and collimating the beam are single aspheric lenses with AR coatings (C240TME-B, Thorlabs).
Furthermore, in the aim of improving the spatial mode matching between the TEM$_{00}$ in free-space and LP$_{01}$ in PM fibers, we developed a special fiber alignment device (FA1000S, FMD Corporation).
As a result, we obtained $92\%$ in the throughput of the entire fiber delay line, which was kept as high as $89\%$ during 6 hours.
Since small temperature changes around the fiber cause drastic changes of the optical pass length resulting in the instability of phase locking,
the fiber is placed inside a box consisting of heat insulating material and vibration-proofing materials.

The Extended EPR states are measured by homodyne detectors.
The two homodyne measurements are performed with balanced beam-splitters and continuous-wave local oscillator (LO) beams. 
To optimise the spatial mode-matching, the LO beams are first filtered by a separate cavity which is a bow-tie cavity consisting of two spherical mirrors (radius of curvature $38$~mm) and two flat mirrors.
Its round trip is $230$~mm length, finesse is $25$, the half width at half maximum is $27$~MHz, and the input and output coupler transmissivities are both $5\%$.
Visibilities are above $98\%$ for every pair of signal beams through free space including the LO beams. The average visibility through the fiber is $95\%$.
The propagation efficiencies from the OPOs to the homodyne detectors are $85$--$96\%$. 
The quantum efficiencies of photodiodes (special order, Hamamatsu Photonics) used in homodyne detectors are about $99\%$, while the bandwidth of the detectors are above $20$~MHz. The LO power is set to $10$~mW for every homodyne measurement.

The signals from the homodyne detectors in the time-domain are stored by an oscilloscope (DPO~7054, Tektronix).
The sampling rate of the oscilloscope is set to $200$~MHz in order to sample enough data points in each wave packet.
Each frame of $2.5$~ms contains $500,000$ points, corresponding to about $30,000$ wave-packets. For each quadrature measurement of each wave-packet we measure $3,000$ frames in order to gather enough statistics to calculate variances.
The quadrature amplitudes are elicited from them by using the temporal mode function $f(t)$.
\begin{align}
\label{modefunction}
\hat \xi_k=\int_{-T/2}^{T/2}  \hat \xi(t-kT)f(t)\mathrm{d}t,
\end{align}
where 
$\hat \xi_k\ (\hat \xi =\hat x, \hat p)$ are discretized quadratures, $\hat \xi(t)\ (\hat \xi =\hat x, \hat p)$ are continuous quadratures from the homodyne detectors, and
$f(t)$ is a Gaussian filter $f(t)\propto \mathrm{e}^{-(\Gamma t)^2}$
which is normalised as $\int_{-T/2}^{T/2}  \left|f(t)\right|^2\mathrm{d}t=1$.
While the bandwidth $\Gamma$ can be chosen arbitrarily, the interval of the wave-packet $T$ depends on the fiber length.
The parameters used here are $T=157.5$~ns and $\Gamma=2\pi \times 2.5$~MHz ($1/\Gamma\sim 64$~ns).

It is necessary to lock the relative phases of beams at every point where the beams interfere.
For this purpose, we utilise bright laser beams of about $10$~$\mu$W, which are directed by homodyne detectors.
However, their laser noise interferes with our measurement results.
To avoid this problem, we switch between data acquisition and feedback control periodically at a switching frequency of $25$~Hz.
Furthermore, to realise strong and reliable phase locking, we implement a custom-made digital feedback control system via field programmable gate arrays (NI PXI-7853R, National Instruments) in PXI chassis (NI PXI-1033, National Instruments).

\subsection{Animation}

\begin{figure}[!tb]
\centering
\includegraphics[scale=0.7,clip]{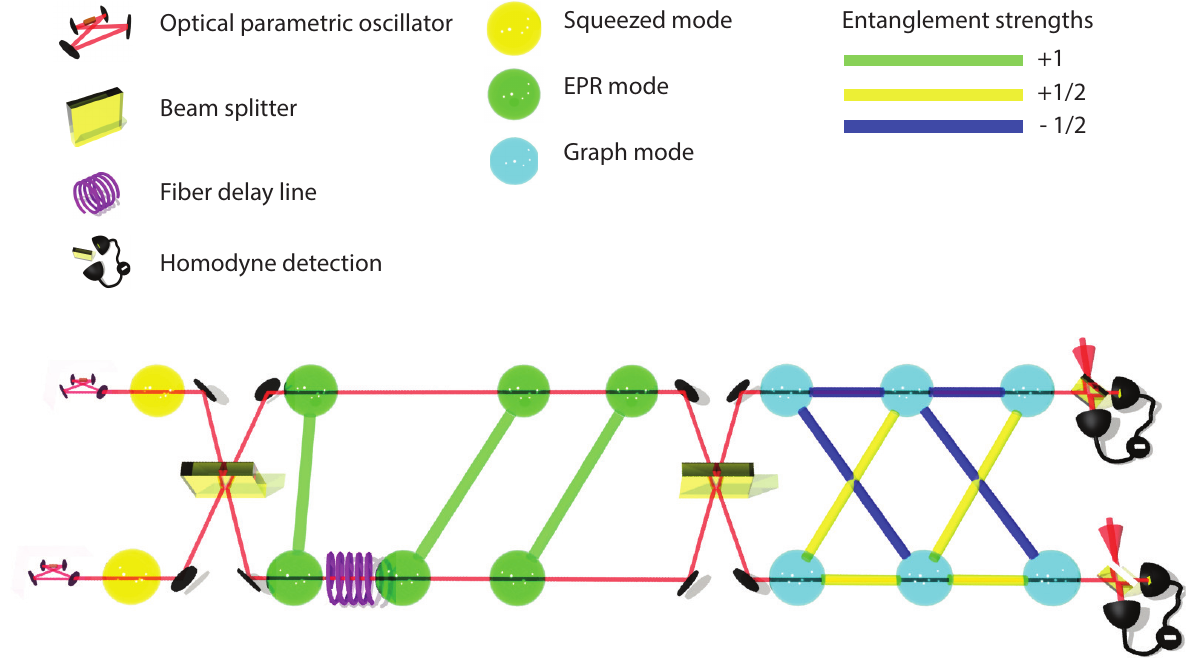}
\caption{Legend for animation of experimental setup.}
\label{fig:AnimationLegend}
\end{figure}

\noindent
An animation can be found at (http://www.alice.t.u-tokyo.ac.jp/Graph-animation.avi) that shows temporal modes propagating through the experimental setup. \figureref{AnimationLegend} is provided as a legend for the animation.

\section{Theory}
\subsection{Derivation of Extended EPR States}

\begin{figure}[!t]
\centering
\includegraphics[scale=1,clip]{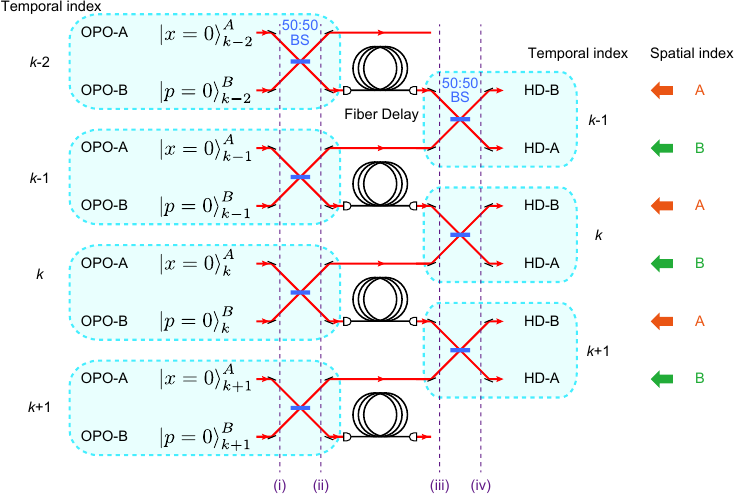}
\caption{The equivalent quantum circuit to the temporal-mode method setup.}
\label{fig:EquivalentQC}
\end{figure}

\noindent
An equivalent linear optics network to our experimental setup is represented in \figref{EquivalentQC}.
In this section we derive the expressions of the extended EPR state by following this circuit with both Schr\"{o}dinger and Heisenberg evolutions.
In the Schr\"{o}dinger picture, we assume the ideal case where the resource squeezing levels are infinite. 
On the other hand we can calculate experimentally realistic expressions in the Heisenberg picture.

\subsubsection{Schr\"{o}dinger Picture in the Ideal Case}
\noindent
Here we utilise infinite squeezing for simplicity.
As per the following calculations with Schr\"{o}\-dinger evolution, the output state is a simultaneous eigenstate of nullifiers.
First, there are position and momentum eigenstates with zero eigenvalue in each temporal location $k$.
Each row in \figref{EquivalentQC} shows the spatial mode index which the temporal-mode method would correspond to. The ket vector $|\mathrm{i}\rangle$ in step (i) is represented as 
\begin{align}
|\mathrm{i}\rangle
&=\prod_k |x=0\rangle^A _k |p=0\rangle^B _k 
\propto \prod_k\int |x=0\rangle^A _k |x=a_{k}\rangle^B _k \,\mathrm{d}a_{k}.
\end{align}
Note that we omit the interval of an integral in this text when the integral interval is from $-\infty$ to $\infty$.
Second, $x$ and $p$ eigenstates in each temporal location are combined by the first beam-splitters.
Here, the beam-splitter operator $\hat B_{i,j}$ for mode $i$ and $j$ is defined as
$\hat B_{i,j}|x=a\rangle_i |x=b\rangle_j=|x=
(a+b)
/\sqrt{2}
\rangle_i |x=
(-a+b)
/\sqrt{2}
\rangle_j$.
Therefore $|\mathrm{ii}\rangle$ becomes
\begin{align}
|\mathrm{ii}\rangle
&=\prod_k\hat B_{A  k, B  k}|\mathrm{i}\rangle
\propto  \prod_k\int \left|x=\tfrac{a_{k}}{\sqrt{2}}\right\rangle^A _k \left|x=\tfrac{a_{k}}{\sqrt{2}}\right\rangle^B _k \mathrm{d}a_{k}
.
\end{align}
Then an optical delay is implemented to spatial mode $B$.
It is equivalent to a delay in the temporal mode index: $k\to k+1$,
\begin{align}
|\mathrm{iii}\rangle
&\propto  \prod_k\int \left|x=\tfrac{a_{k}}{\sqrt{2}}\right\rangle^A _k \left|x=\tfrac{a_{k}}{\sqrt{2}}\right\rangle^B _{k+1} \mathrm{d}a_{k}
=\int \prod_k \left|x=\tfrac{a_{k}}{\sqrt{2}}\right\rangle^A _k \left|x=\tfrac{a_{k-1}}{\sqrt{2}}\right\rangle^B _{k} \mathrm{d}a_k
.
\end{align}
Finally, the beams are combined on the second beam-splitters, giving
\begin{align}
|\mathrm{iv}\rangle
&=\prod_k \hat B_{B  k,A  k}|\mathrm{iii}\rangle
\propto \int \prod_k \left|x=\tfrac{1}{2}(a_k-a_{k-1})\right\rangle^A _k \left|x=\tfrac{1}{2}(a_k+a_{k-1})\right\rangle^B _{k} \mathrm{d}a_k.
\end{align}
In the same manner the expressions in the $p$ basis can also be calculated as 
\begin{align}
|\mathrm{iv}\rangle
&\propto \int \prod_k \left|p=\tfrac{1}{2}(b_k+b_{k-1})\right\rangle^A _k \left|p=\tfrac{1}{2}(b_k-b_{k-1})\right\rangle^B _{k} \mathrm{d}b_k
.
\end{align}
Since the output extended EPR state $\left| \mathrm{XEPR}\right\rangle$ is equal to $|\mathrm{iv}\rangle$, the nullifiers obviously become zero as
\begin{align}
\left( \hat x^A_k+\hat x^B_k+\hat x^A_{k+1}-\hat x^B_{k+1}\right) \left| \mathrm{XEPR}\right\rangle
& =0, \\
\left( \hat p^A_k+\hat p^B_k-\hat p^A_{k+1}+\hat p^B_{k+1}\right) \left| \mathrm{XEPR}\right\rangle
& =0. 
\end{align}

\subsubsection{Heisenberg Evolution with Finite Squeezing}
\noindent
In the Heisenberg evolution, the variances of nullifiers in the case of finite resource squeezing levels can be calculated.
The complex amplitudes $\hat a^{A \mathrm{(i)}}_k$ and $\hat a^{B \mathrm{(i)}}_k$ of the initial state in step (i) are represented as
\fleqnon
\begin{align}
\mathrm{(i)}&\qquad
	\hat a^{A \mathrm{(i)}}_k=\mathrm{e}^{-r_A }\hat x^{A \mathrm{(0)}}_k+i\,\mathrm{e}^{r_A }\hat p^{A \mathrm{(0)}}_k,\qquad 
	\hat a^{B \mathrm{(i)}}_k=\mathrm{e}^{r_B }\hat x^{B \mathrm{(0)}}_k+i\,\mathrm{e}^{-r_B }\hat p^{B \mathrm{(0)}}_k,
\end{align}
\fleqnoff
where $\mathrm{e}^{-r_A }\hat x^{A \mathrm{(0)}}_k$ and $\mathrm{e}^{-r_B }\hat p^{B \mathrm{(0)}}_k$ are the squeezed quadratures of the $k$-th squeezed state in the spatial location $A$ and $B$, respectively. So we have
\begin{align}
&\Expbig{\hat x_k^{A(0)}}=
\Expbig{\hat p_k^{A(0)}}=
\Expbig{\hat x_k^{B(0)}}=
\Expbig{\hat p_k^{B(0)}}=0,
\\
&\Expbig{\bigl(\hat x_k^{A(0)}\bigr)^2}=
\Expbig{\bigl(\hat p_k^{A(0)}\bigr)^2}=
\Expbig{\bigl(\hat x_k^{B(0)}\bigr)^2}=
\Expbig{\bigl(\hat p_k^{B(0)}\bigr)^2}=\frac{1}{4}.
\end{align}
After combining the terms through the action of a beam-splitter they become
\fleqnon
\begin{align}
\mathrm{(ii)}&\qquad
\begin{pmatrix}
\hat a^{A \mathrm{(ii)}}_k \\
\hat a^{B \mathrm{(ii)}}_k
\end{pmatrix}
=\hat B_{A  k, B  k}^\dagger 
\begin{pmatrix}
\hat a^{A \mathrm{(i)}}_k \\
\hat a^{B \mathrm{(i)}}_k
\end{pmatrix}
\hat B_{A  k, B  k}
=\frac{1}{\sqrt{2}}
\begin{pmatrix}
1 & 1 \\
-1 & 1
\end{pmatrix}
\begin{pmatrix}
\hat a^{A \mathrm{(i)}}_k \\
\hat a^{B \mathrm{(i)}}_k
\end{pmatrix}.
\end{align}
Subsequently, the of optical delay for mode $B$ is represented as
\begin{align}
\mathrm{(iii)}&\qquad
\hat a^{A \mathrm{(iii)}}_k=\hat a^{A \mathrm{(ii)}}_k,
\qquad
\hat a^{B \mathrm{(iii)}}_k=\hat a^{B \mathrm{(ii)}}_{k-1}.
\end{align}
Finally, by combining them on the last beam-splitter the final complex amplitudes of the output state are given:
\begin{align}
\mathrm{(iv)}&\qquad
\begin{pmatrix}
\hat a^{B \mathrm{(iv)}}_k \\
\hat a^{A \mathrm{(iv)}}_k 
\end{pmatrix}
=\hat B_{B  k,A  k}^\dagger 
\begin{pmatrix}
\hat a^{B \mathrm{(iii)}}_k \\
\hat a^{A \mathrm{(iii)}}_k 
\end{pmatrix}
\hat B_{B  k,A  k}
=\frac{1}{\sqrt{2}}
\begin{pmatrix}
1 & 1 \\
-1 & 1
\end{pmatrix}
\begin{pmatrix}
\hat a^{B \mathrm{(iii)}}_k \\
\hat a^{A \mathrm{(iii)}}_k 
\end{pmatrix}
\notag \\
&\hspace{6.2em}
=
\frac{1}{2}
\begin{pmatrix}
\hat a^{A \mathrm{(i)}}_k
+\hat a^{B \mathrm{(i)}}_k
-\hat a^{A \mathrm{(i)}}_{k-1}
+\hat a^{B \mathrm{(i)}}_{k-1}
\\
\hat a^{A \mathrm{(i)}}_k
+\hat a^{B \mathrm{(i)}}_k
+\hat a^{A \mathrm{(i)}}_{k-1}
-\hat a^{B \mathrm{(i)}}_{k-1}
\end{pmatrix},
\end{align}
\fleqnoff
Since
$\hat a^A_k=\hat a^{A \mathrm{(iv)}}_k$
and
$\hat a^B_k=\hat a^{B \mathrm{(iv)}}_k $,
the ideal nullifiers of the extended EPR state are expressed as
\begin{align}
 \hat x^A_k+\hat x^B_k+\hat x^A_{k+1}-\hat x^B_{k+1}
&=2\, \mathrm{e}^{-r_A } \hat x_k^{A  (0)}, \\
\hat p^A_k+\hat p^B_k-\hat p^A_{k+1}+\hat p^B_{k+1}
&=2\, \mathrm{e}^{-r_B } \hat p_k^{B  (0)}.
\end{align}
Therefore, we can calculate the nullifier variances which determine the theoretical value of the inseparable condition as shown in the main text [Eq.~(3)], 
\begin{align}
\label{nullifier}
\bigl\langle \left( \hat x^A_k+\hat x^B_k+\hat x^A_{k+1}-\hat x^B_{k+1}\right)^2\bigr\rangle 
&=\mathrm{e}^{-2r_A }
<\frac{1}{2},\\
\mathrm{and} \quad \bigl\langle \left( \hat p^A_k+\hat p^B_k-\hat p^A_{k+1}+\hat p^B_{k+1}\right)^2\bigr\rangle 
&=\mathrm{e}^{-2r_B }
<\frac{1}{2}.
\end{align}
This shows that the sufficient condition for inseparability is satisfied when $-3$~dB resource squeezing in each OPO is available.

\subsection{Inseparability Criteria for Extended EPR States}

\noindent
Here, we discuss sufficient conditions of entanglement for the extended EPR states, 
based on the van Loock-Furusawa inseparability criteria \cite{Peter03}.
We consider all of the cases where an approximate extended EPR state is separable into two subsystems $S_1$ and $S_2$. 
A necessary condition of separability is obtained as an inequality for each case. 
If all of the separable cases are denied by not satisfying the inequalities, the state is proved to be in an entangled state with full inseparability. 

First, we consider the combinations of four nodes $\{ A_k, B_k, A_{k+1}, B_{k+1}\}$ distributed into the two subsystems.
When all of the four are not distributed into either subsystem, the possible cases are as below.
Here we abbreviate the nullifiers as 
$\hat X_k\equiv \hat x^A_k+\hat x^B_k+\hat x^A_{k+1}-\hat x^B_{k+1}$
and
$\hat P_k\equiv \hat p^A_k+\hat p^B_k-\hat p^A_{k+1}+\hat p^B_{k+1}$.

\begin{itemize}
\item Case: $\{A_k \} \subset S_1$, $\{B_k,A_{k+1},B_{k+1}\} \subset S_2$.

The addition of nullifier variances always satisfies the following inequality:
\begin{align}
\label{first}
\Varbig{\hat X_k}+\Varbig{\hat P_k}\geq\frac{1}{2}\left( |1| +|1-1-1|\right) =1.
\end{align}

\item Case: $\{B_k \} \subset S_1$, $\{A_k,B_k,B_{k+1}\} \subset S_2$.

The addition of nullifier variances always satisfies the following inequality:
\begin{align}
\Varbig{\hat X_k}+\Varbig{\hat P_k}\geq\frac{1}{2}\left( |1| +|1-1-1|\right) =1.
\end{align}

\item Case: $\{A_{k+1} \} \subset S_1$, $\{A_k,B_k,B_{k+1}\} \subset S_2$.

The addition of nullifier variances always satisfies the following inequality:
\begin{align}
\Varbig{\hat X_k}+\Varbig{\hat P_k}\geq\frac{1}{2}\left( |-1| +|1+1-1|\right) =1.
\end{align}

\item Case: $\{B_{k+1} \} \subset S_1$, $\{A_k,B_k,A_{k+1}\} \subset S_2$.

The addition of nullifier variances always satisfies the following inequality:
\begin{align}
\Varbig{\hat X_k}+\Varbig{\hat P_k}\geq\frac{1}{2}\left( |-1| +|1+1-1|\right) =1.
\end{align}

\item Case: $\{A_k,B_k \} \subset S_1$, $\{A_{k+1},B_{k+1}\} \subset S_2$.

The addition of nullifier variances always satisfies the following inequality:
\begin{align}
\Varbig{\hat X_k}+\Varbig{\hat P_k}\geq\frac{1}{2}\left( |1+1| +|-1-1|\right) =2.
\end{align}

\item Case: $\{A_k,A_{k+1}\} \subset S_1$, $\{B_k,B_{k+1} \} \subset S_2$.

The addition of nullifier variances always satisfies the following inequality:
\begin{align}
\Varbig{\hat X_k}+\Varbig{\hat P_{k+1}}\geq\frac{1}{2}\left( |1| +|-1|\right) =1.
\end{align}

\item Case: $\{A_k,B_{k+1} \} \subset S_1$, $\{B_k,A_{k+1}\} \subset S_2$.

The addition of nullifier variances always satisfies the following inequality:
\begin{align}
\label{last}
\Varbig{\hat X_k}+\Varbig{\hat P_{k+1}}\geq\frac{1}{2}\left( |-1| +|1|\right) =1.
\end{align}

\end{itemize}

Therefore, when the inequalities
$\Varbig{\hat X_k}+\Varbig{\hat P_{k}}<1
$
and
$\Varbig{\hat X_k}+\Varbig{\hat P_{k+1}}<1
$
are satisfied,
any of the seven inequalities \eqref{first}--\eqref{last} is not satisfied, 
which means that the four nodes $\{ A_k, B_k, A_{k+1}, B_{k+1}\}$ are not separable into two subsystem $S_1$ and $S_2$.

Then, we apply the same discussion for all temporal indices $k$.
When the inequalities 
$\Varbig{\hat X_k}+\Varbig{\hat P_{k}}<1
$
and
$\Varbig{\hat X_k}+\Varbig{\hat P_{k+1}}<1
$
are satisfied for all $k$, any partitioning of the whole system is denied, 
which means that all nodes are entangled.
We may take a more severe but simpler sufficient condition for entanglement as
\begin{align}
\Varbig{\hat X_k}<\frac{1}{2}
\quad
\mathrm{and}
\quad
\Varbig{\hat P_k}<\frac{1}{2},
\end{align}
for all $k$,
which is shown in Eq.~(3) of the main text.


\subsection{Graph Correspondence}
\noindent
In this section, we discuss the intuitive representation of the extended EPR state in terms of the graphical calculus for Gaussian pure states~\cite{Menicucci2011}. Every $N$-mode zero-mean Gaussian pure state can be uniquely represented by an undirected complex-weighted graph~$\mat Z$, whose imaginary part (i.e., that of the adjacency matrix for the graph) is positive definite. (In what follows, we make no distinction between a graph and its adjacency matrix.) The graph~$\mat Z$ shows up directly in the position-space wavefunction for the corresponding state~$\ket {\psi_{\mat Z}}$ (with $\hbar = \tfrac 1 2$):
\begin{align}
	\lsub {\inprod {\vec s} {\psi_{\mat Z}}} {x} = \psi_{\mat Z}(\vec s) \propto \exp\left[ i \vec s^\tp \mat Z \vec s \right].
\end{align}
\setcitestyle{numbers,open={},close={},citesep={,}}
Any Gaussian pure state~$\ket{\psi_{\mat Z}}$ satisfies a set of exact nullifier relations based on its associated complex matrix~$\mat Z$ (ref.~\cite{Menicucci2011}):
\setcitestyle{super,open={},close={},citesep={,}}
\begin{align}
 \label{eq:graph_rep}
	\left(\opvec p - \mat Z \opvec x \right) \ket {\psi_{\mat Z}} = \vec 0,
\end{align}
where $\opvec p$ and $\opvec x$ are column vectors of momentum and position operators, respectively. The special case of the $N$-mode ground state ($\mat Z_{\text{ground}} = i\mat \id$) is easy to verify by noting that the vector of nullifiers in that case is just the vector of annihilation operators.

\setcitestyle{numbers,open={},close={},citesep={,}}
The extended EPR state is exactly the state originally proposed by Menicucci in ref.~\cite{Menicucci11}.\setcitestyle{super,open={},close={},citesep={,}}In that proposal, it was shown that such a state is locally equivalent (up to phase shifts on half the modes) to a CV cluster state, which is a universal resource for measurement-based quantum computing with continuous variables~\cite{Menicucci06, Gu2009}. Since measurement-based quantum computation requires the ability to do homodyne detection of any (rotated) quadrature, plus photon counting~\cite{Gu2009}, the phase shifts required to transform the generated state (the extended EPR state) into a CV cluster state do not need to be physically performed on the state after generation. Instead, one can account for them entirely just by updating the measurement-based protocol to be implemented (i.e., redefine quadratures $\op x \to \op p$ and $\op p \to -\op x$ on the appropriate modes)~\cite{Menicucci11}. Because of this equivalence, the original proposal~\cite{Menicucci11} used a simplified graphical calculus that blurred the distinction between the extended EPR state and corresponding CV cluster state since the two were, for quantum computational purposes, effectively the same resource. The distinction between these states turns out to make a huge difference, however, when one tries to experimentally characterize the generated state. In this case, it is much easier to work with the mathematics of the extended EPR state.

\setcitestyle{numbers,open={},close={},citesep={,}}
For clarity and completeness, here we present the full complex-weighted graph~$\mat Z_E$ (ref.~\cite{Menicucci2011}) corresponding to the extended EPR state originally proposed in ref.~\cite{Menicucci11} and reported on in this work:
\setcitestyle{super,open={},close={},citesep={,}}
\begin{figure}[H]
\centering
\includegraphics[width=.9\textwidth,clip]{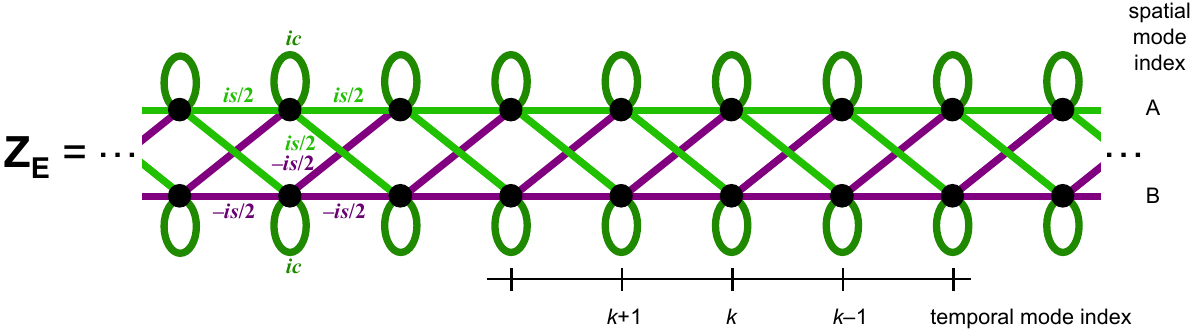}
\caption{The full graph~\cite{Menicucci2011} for the extended EPR state.
}
\label{fig:GraphZE}
\end{figure}
\setcitestyle{numbers,open={},close={},citesep={,}}
\noindent In this graph, $c = \cosh 2r$ and $s = \sinh 2r$, with $r$ being the initial squeezing parameter of the states emitted by the OPO. (See ref.~\cite{Menicucci2011} for more details on such graphs.) The green (purple) edges have positive- \mbox{(negative-)imaginary} weight $\pm \tfrac i 2 \sinh 2r$, and the green self-loops have positive-imaginary weight $i\cosh 2r$. 
This state can be transformed, by $-\tfrac \pi 2$ phase shifts on both modes of all odd (or all even) time indices, into the following CV cluster state~$\mat Z_C$ (ref.~\cite{Menicucci11}):
\setcitestyle{super,open={},close={},citesep={,}}
\begin{figure}[H]
\centering
\includegraphics[width=.9\textwidth,clip]{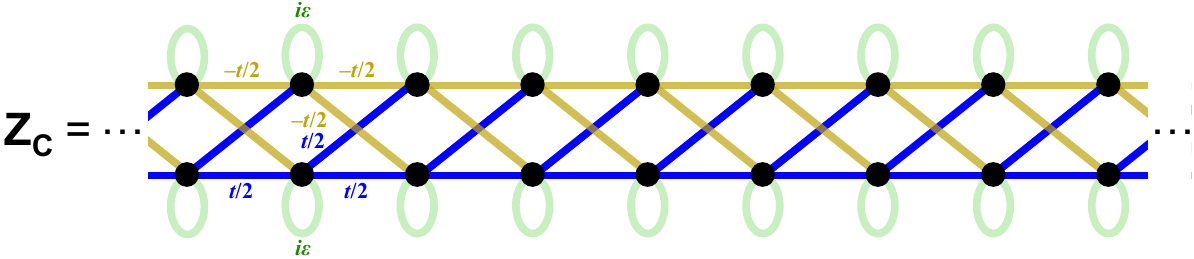}
\caption{The full graph~\cite{Menicucci2011} for the CV cluster state obtained by phase shifting both spatial modes in every other time index of the extended EPR state.
}
\label{fig:GraphZC}
\end{figure}
\noindent In this graph, $t = \tanh 2r$ and $\varepsilon = \sech 2r$. The blue (yellow) edges have positive- \mbox{(negative-)real} weight $\pm \tfrac 1 2 \tanh 2r$, and the green self-loops have positive-imaginary weight $i\cosh 2r$. Darker colours indicate larger magnitude of the corresponding edge weight. In the large-squeezing limit, $t = \tanh 2r \to 1$, and $\varepsilon = \sech 2r \to 0$, which allows us to define an unphysical, ideal CV cluster-state graph~$\mat G$ to which $\mat Z_C$ is a physical approximation:
\begin{figure}[H]
\centering
\includegraphics[width=.9\textwidth,clip]{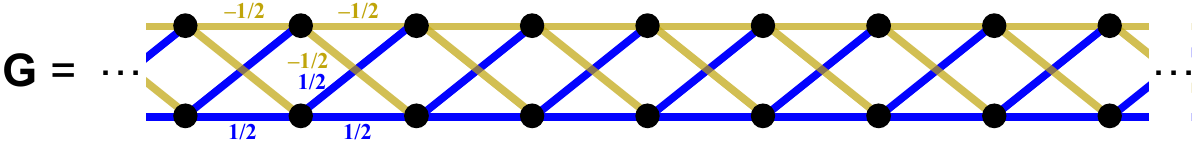}
\caption{The ideal CV cluster-state graph in the limit $r \to \infty$.}
\label{fig:GraphG}
\end{figure}
\noindent Notice that
\begin{align}
\label{eq:ZEfromG}
	\mat Z_E &= i (\cosh 2r) \mat \id - i (\sinh 2r) \mat G, \\
\label{eq:ZCfromG}
 	\mat Z_C &=  i (\sech 2r) \mat \id + (\tanh 2r) \mat G. 
\end{align}
\noindent The crucial properties of $\mat G$ that enable such a simple connection between $\mat Z_E$, $\mat Z_C$, and $\mat G$ are (a) that $\mat G$ is bipartite and (b) that $\mat G$ is self-inverse (i.e., $\mat G^2 = \mat \id$ as a matrix). These mathematical properties allow all three of these graphs to be visually similar.

With this simplification in hand, we can derive new nullifier relations for $\mat Z_E$ in terms of $\mat G$. We start by observing that premultiplying both sides of Eq.~\eqref{eq:graph_rep} by $-\mat Z^{-1}$ gives the additional exact nullifier relation
\begin{align}
	\left(\opvec x - \mat Z^{-1} \opvec p \right) \ket {\psi_{\mat Z}} = \vec 0.
\end{align}
Substituting $\mat Z \to \mat Z_E$ [Eq.~\eqref{eq:ZEfromG}] and noting that $\mat Z_E^{-1} = -i (\cosh 2r) \mat \id - i (\sinh 2r) \mat G$, we have the two exact nullifier relations
\begin{align}
	\left[\opvec p - (i c \mat \id - i s \mat G) \opvec x \right] \ket {\psi_{\mat Z_{E}}} &= \vec 0, \nonumber \\
	\left[\opvec x + (i c \mat \id + i s \mat G) \opvec p \right] \ket {\psi_{\mat Z_{E}}} &= \vec 0.
\end{align}
By premultiplying by $\pm i \varepsilon$, respectively, we obtain
\begin{align}
	\left[i\varepsilon \opvec p + ( \opvec x - t \mat G \opvec x) \right] \ket {\psi_{\mat Z_{E}}} &= \vec 0, \nonumber \\
	\left[-i\varepsilon \opvec x + ( \opvec p + t \mat G \opvec p) \right] \ket {\psi_{\mat Z_{E}}} &= \vec 0.
\end{align}
In the large-squeezing limit ($r \to \infty$), $\varepsilon \to 0$ and $t \to 1$, and we have the following approximate nullifiers for the extended EPR state:
\begin{align}
	(\opvec x - \mat G \opvec x) \ket {\psi_{\mat Z_{E}}} \xrightarrow{r \to \infty} \vec 0, \nonumber \\
	(\opvec p + \mat G \opvec p) \ket {\psi_{\mat Z_{E}}} \xrightarrow{r \to \infty} \vec 0.
\end{align}
This is the state we have created. For completeness, however, we can compare these to the exact and approximate nullifiers for the associated CV cluster state~$\ket{\psi_{\mat Z_C}}$ obtained by phase shifting particular nodes as described above (either actively or by simply redefining quadratures used for the measurements). Substituting $\mat Z \to \mat Z_C$ [Eq.~\eqref{eq:ZCfromG}] and noting that $\mat Z_C^{-1} = -i (\sech 2r) \mat \id + (\tanh 2r) \mat G$, the exact nullifiers are
\begin{align}
	\left(- i \varepsilon \opvec x + \opvec p - t \mat G \opvec x \right) \ket {\psi_{\mat Z_C}} = \vec 0, \nonumber \\
	\left(i \varepsilon \opvec p + \opvec x - t \mat G \opvec p \right) \ket {\psi_{\mat Z_C}} = \vec 0,
\end{align}
which, in the large-squeezing limit, reduce to the following approximate nullifiers:
\begin{align}
	(\opvec p - \mat G \opvec x) \ket {\psi_{\mat Z_{C}}} \xrightarrow{r \to \infty} \vec 0, \nonumber \\
	(\opvec x - \mat G \opvec p) \ket {\psi_{\mat Z_{C}}} \xrightarrow{r \to \infty} \vec 0.
\end{align}
Once again, these simple and symmetric expressions in terms of $\mat G$ are unusual and only possible because $\mat G$ is bipartite and self-inverse~\cite{Menicucci2011}.

\subsection{Equivalence to Sequential Teleportation-based Quantum Computation Circuit}
\setcitestyle{numbers,open={},close={},citesep={,}}
\noindent
In reference \cite{Menicucci11}, Menicucci proposed that by erasing half of the state (one rail), the cluster states can be used as resources for measurement-based quantum computation (MBQC). 
However, erasing half of the state is a wasteful process and it is experimentally hard to perform the necessary feedforwards to future and past modes in the time axis. 
Here, we show that the extended EPR state is a resource for MBQC, and that we can fully utilise every degree of freedom without erasing half of the state.
We devise a much more efficient method of using this resource state for quantum computation than the method originally proposed in ref.~\cite{Menicucci11}, in terms of its use of the available squeezing resources.\setcitestyle{super,open={},close={},citesep={,}}Specifically, arbitrary Gaussian operation may be implemented by only $4$ measurements, which is more efficient than the 8 measurements necessary with the original method \cite{Alexander13}.
Furthermore, we show that non-Gaussian operations may be performed on the extended EPR state by introducing non-Gaussian measurements, leading to one-mode universal MBQC.

\subsubsection{Gaussian Operation}

\begin{figure}[!b]
\centering
\includegraphics[scale=1,clip]{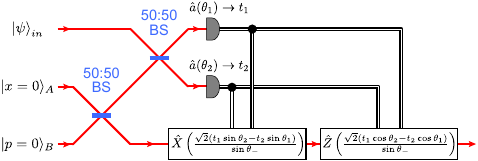}
\caption{Teleportation-based circuit}
\label{fig:teleportation}
\end{figure}

\noindent
First of all,
let us consider the quantum teleportation-based circuit shown in \figref{teleportation}.
The resource EPR state is generated by combining position and momentum eigenstates via the first beam-splitter.
After the input state $|\psi\rangle$ is coupled via a second beam-splitter, 
two observables 
$\hat a_{in}(\theta_1)$ and $\hat a_A(\theta_2)$
$(\hat a(\theta)=\hat x\cos\theta +\hat p\sin\theta)$
are measured, 
giving the measurement results $t_1$ and $t_2$.
Then, the corresponding feedforward operations
$\hat X\bigl(
\sqrt{2}
(
t_1\sin\theta_2
-t_2\sin\theta_1
)
/
\sin \theta_-
\bigr)
$
and
$\hat Z\bigl(
\sqrt{2}
(
t_1\cos\theta_2
-t_2\cos\theta_1
)
/\sin \theta_-
\bigr)
$
are performed on the remaining mode,
where 
$\hat X(s)=\mathrm{e}^{-2is\hat p}$ and $\hat Z(s)=\mathrm{e}^{2is\hat x}$ are the position and momentum displacement operators,
and $\theta_\pm$ is $\theta_{\pm}=\theta_1 \pm \theta_2$.
The resulting ket vector $|out\rangle$ becomes
\begin{align}
|out\rangle=
\hat R \left(-\theta_+/2+\pi/2\right)
\hat S\left( \log \tan(\theta_-/2)\right)
\hat R \left(-\theta_+/2\right)
|\psi\rangle.
\end{align}
Here, $\hat R (\theta) =\mathrm{e}^{i\theta (\hat x^2 +\hat p^2)}$ and $\hat S(r)=\mathrm{e}^{ir(\hat x\hat p+\hat p \hat x)}$ are the $\theta$ rotation operator and squeezing operator, respectively, in phase space.

\begin{figure}[!tb]
\centering
\includegraphics[scale=.88,clip]{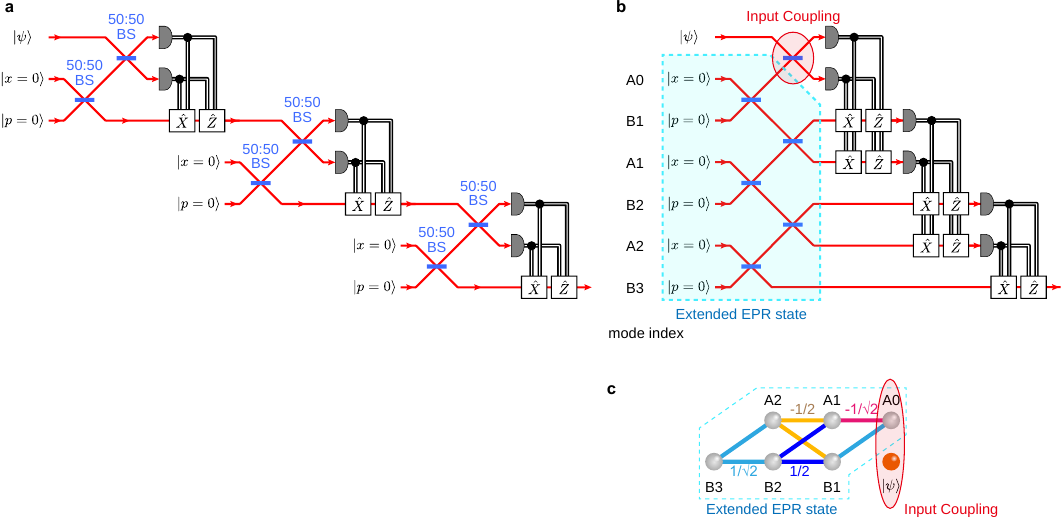}
\caption{The resource for MBQC with extended EPR states. \textbf{a}, Sequential teleportation circuit. \textbf{b}, Equivalent circuit. \textbf{c}, Abstract illustration of sequential teleportation with an extended EPR state.}
\label{fig:SequentialTeleportation}
\end{figure}

\begin{figure}[!tb]
\centering
\includegraphics[scale=.93,clip]{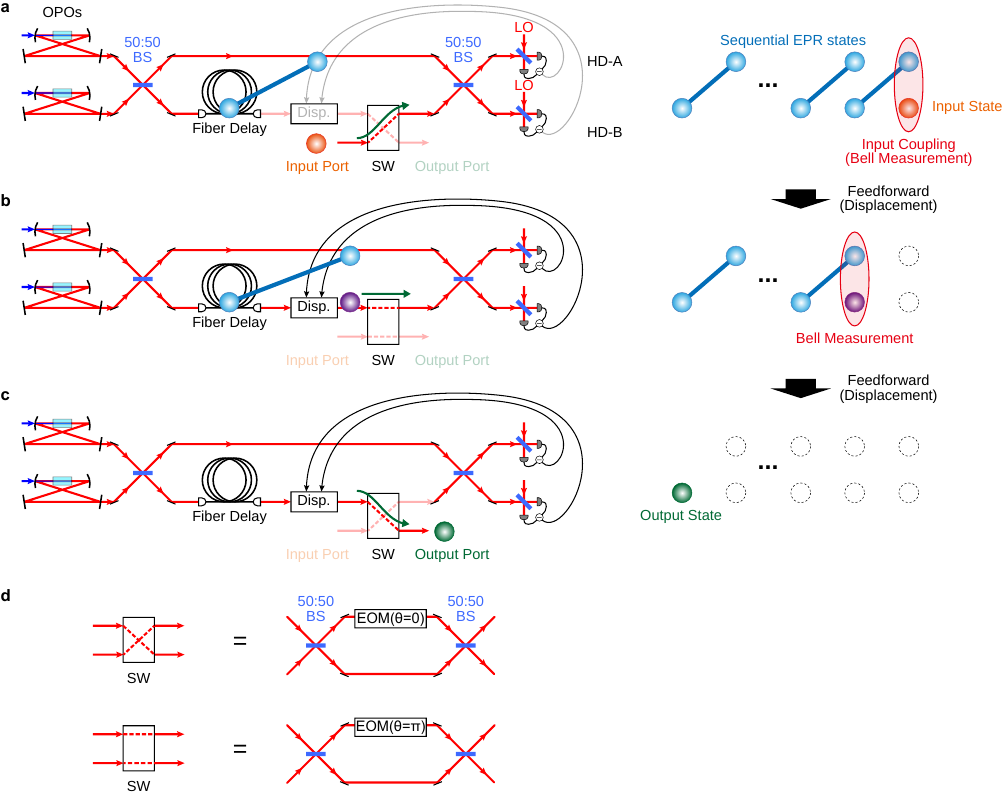}
\caption{Experimental setup (left) and abstract illustration (right) for MBQC with extended EPR states.
\textbf{a}, Input coupling.
\textbf{b}, Sequential MBQC.
\textbf{c}, Output.
\textbf{d}, Switching (SW) device with Mach-Zehnder interferometer including electro optical modulator (EOM).
}
\label{fig:SequentialTeleportationSetup}
\end{figure}

We may combine this teleportation-based circuit sequentially as shown in \figref{SequentialTeleportation}a.
It can be experimentally realised by modifying our experimental setup as shown in \figref{SequentialTeleportationSetup},
where the input-output coupling port is realised by a switching device to a Mach-Zehnder interferometer containing a configurable phase-shifter.
In \figref{SequentialTeleportation}a, the input-coupling beam-splitters and displacement operators can be exchanged as
$\hat B_{i,j} \hat X_i (t)=\hat X_i (t/\sqrt{2})\hat X_j (-t/\sqrt{2})\hat B_{i,j}$
and
$\hat B_{i,j} \hat Z_i (t)=\hat Z_i (t/\sqrt{2})\hat Z_j (-t/\sqrt{2})\hat B_{i,j}$.
As a result, an equivalent circuit is \figref{SequentialTeleportation}b.
Here, in the region enclosed by the dotted line, is the part of the circuit that creates the extended EPR state, which is represented as a graph as shown in \figref{SequentialTeleportation}c.
This shows that the extended EPR state can be used as a resource for MBQC.
Therefore, by adding input-coupling, measurement and feedforward optics to our setup, it will become a MBQC circuit.

\subsubsection{Non-Gaussian Operation}

\begin{figure}[!t]
\centering
\includegraphics[scale=1,clip]{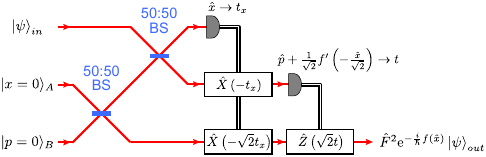}
\caption{Teleportation-based non-Gaussian operation circuit}
\label{fig:NonGaussian}
\end{figure}

\noindent
Non-Gaussian operations may also be implemented by using the teleportation-based circuit shown in \figref{NonGaussian}.
It is derived in the following way.
The ket vector after input coupling is represented as 
\begin{align}
\notag
\hat B&_{in,A}|\psi\rangle_{in}
\otimes\left(
	\hat B_{A,B} |x=0\rangle_A |p=0\rangle_B \right)\\
	&\propto
		\int \mathrm{d}t_x|x=t_x\rangle_{in}\hat X_A(t_x)\hat X_B(\sqrt{2}\,t_x)
		\otimes
		\left( \int \mathrm{d} \xi_{}\,\psi(\xi_{}) \bigl|x=-\sqrt{2}\,\xi_{}\bigr\rangle_A\bigl|x=-\xi_{}\bigr\rangle_B
		\right).
\end{align}
Therefore, by measuring $\hat x$ on the input mode giving measurement results $t_x$, and performing the displacement operation $\hat X_A(-t_x)$ and $\hat X_B(-\sqrt{2}\,t_x)$,
the resulting state becomes $\int \mathrm{d} \xi_{}\,\psi(\xi_{}) \bigl|x=-\sqrt{2}\,\xi_{}\bigr\rangle_A\bigl|x=-\xi_{}\bigr\rangle_B$.
Further, it is also expressed as
\begin{align}
\notag
\int \mathrm{d} &\xi_{}\,\psi(\xi_{}) \bigl|x=-\sqrt{2}\,\xi_{}\bigr\rangle_A\bigl|x=-\xi_{}\bigr\rangle_B
\\
		&\propto
		\int \mathrm{d}t\left|p+\tfrac{1}{\sqrt{2}}f'\left(-\tfrac{x}{\sqrt{2}}\right)=t\right\rangle_A \hat Z_B(-\sqrt{2}\,t)
			\otimes
		\left(
						\hat F_B^2 \exp\left[-\tfrac{i}{\hbar}f(\hat x_B)\right]|\psi\rangle_B
		\right),
\end{align}
where $f(x)$ is the arbitrary function of $x$ and $\hat F=\hat R(\pi/2)$ is the Fourier transform operator.
In the same way, by measuring $p+f'(-x/\sqrt{2})/\sqrt{2}$ on mode $A$, giving measurement results $t$, and performing the displacement operation $\hat Z_B(\sqrt{2}\,t)$,
the output state becomes $\hat F_B^2 \exp\left[-\tfrac{i}{\hbar}f(\hat x_B)\right]|\psi\rangle_B$.
When $f(x)$ is higher than a quadratic polynomial, it is a non-Gaussian operation.
Note that since it can be accomplished by only using displacement feedforwards, input coupling beam-splitters can also be exchanged.
Therefore, by using the extended EPR state, non-Gaussian operations can be performed sequentially, resulting in a resource for universal one-mode universal MBQC.

\section{Data Analysis}
\subsection{Influence of Experimental Losses}
\noindent
Experimental imperfections lead to degraded resource squeezing levels.
In particular, the unbalanced losses between the optical fiber and free space channels cause the degradation of nullifier variances.
Taking into account these losses, instead of equation \eqref{nullifier} we get more involved theoretical values given by:
\begin{align}
\begin{array}{r}
\bigl\langle \hat X_k^2\bigr\rangle 
=\frac{1}{4}\left(\eta^A  +\eta^A _F\right)^2\left( Sq_A-1\right)
	+\frac{1}{4}\bigl(\eta^B  -\eta^B _F\bigr)^2\left( ASq_B-1\right)
	+1
,\\
\bigl\langle \hat P_k^2\bigr\rangle 
=\frac{1}{4}\bigl(\eta^B  +\eta^B _F\bigr)^2\left( Sq_B-1\right)
	+\frac{1}{4}\left(\eta^A  -\eta^A _F\right)^2\left( ASq_A-1\right)
	+1
,
\end{array}
\end{align}
where $Sq_\Gamma$ and $ASq_\Gamma$ are squeezing and anti-squeezing terms for beam $\Gamma(=A,B)$, and $\eta_\Gamma^2$ and $(\eta_\Gamma^F)^2$ are the effective efficiencies of squeezing levels for beam $\Gamma$ through the free space and optical fiber channels, respectively.
To be more precise, 
$\eta$ is given by: (quantum efficiency at homodyne detector) $\times$ (influence of intracavity loss $T/(T+L)$) $\times$ (visibility between probe and LO beams)$^2$ $\times$ (propagation efficiency),
where $T$ is the transmittance of the output coupler and $L$ is the intracavity loss in the OPO. 
In the experiment, these efficiencies are 
$\eta_A^2=88.2\%$,
$\eta_B^2=89.9\%$,
$(\eta_A^F)^2=73.7\%$
and
$(\eta_B^F)^2=75.3\%$.
Squeezing levels are calculated as 
\begin{align}
Sq= \int  \left|f(\omega)\right| ^2R_- (\omega)\mathrm{d}\omega,
\quad 
ASq= \int  \left|f(\omega)\right| ^2R_+ (\omega)\mathrm{d}\omega,
\\
R_\pm (\omega)=1\pm (1-\eta(\omega)) \frac{4x}{(1\mp x)^2+(\omega/\gamma)^2}
,
\end{align}
where 
$f(\omega)$ is the Fourier transformation of mode function $f(t)$,
$\eta(\omega)$ is the ratio of electrical noise to shot noise at angular frequency $\omega$,
$x$ is the pump parameter which is related to the classical parametric amplification gain $G$ as $G=(1-x)^{-2}$, and
$\gamma$ is the angular frequency half width at half maximum of the OPO.
By substituting in these experimental values,  
we get 
$\bigl\langle \hat X_k^2\bigr\rangle =-5.13$~dB
and
$\bigl\langle \hat P_k^2\bigr\rangle =-5.33$~dB.
They agree well with experimental results.




\paragraph*{Acknowledgments:}
N.C.M.\ is grateful to R.~Alexander, and P.~van~Loock for helpful discussions. \blk

\end{document}